\documentclass[aps,preprint, nofootinbib,floatfix,superscriptaddress]{revtex4}
\usepackage{epsfig}
\usepackage{graphicx}
\usepackage{amsmath,stackrel}
\usepackage{bm}
\usepackage{color, soul}
\usepackage{amsmath}
\usepackage{xcolor}
\usepackage{float}
\def\nn {\nonumber}
\newcommand{\be}{\begin{equation}}
\newcommand{\ee}{\end{equation}}
\newcommand{\bea}{\begin{eqnarray}}
\newcommand{\eea}{\end{eqnarray}}

\newcommand\varpm{\mathbin{\vcenter{\hbox{%
\oalign{\hfil$\scriptstyle+$\hfil\cr
\noalign{\kern-.3ex}
$\scriptscriptstyle({-})$\cr}%
}}}}
\newcommand\varmp{\mathbin{\vcenter{\hbox{%
\oalign{$\scriptstyle({+})$\cr
\noalign{\kern-.3ex}
\hfil$\scriptscriptstyle-$\hfil\cr}%
}}}}

\newcommand{\bx}{\bm x}

\newcommand{\br}{\bm r}
\newcommand{\bk}{\bm k}
\newcommand{\bp}{\bm p}

\newcommand{\bP}{\bm P}

\newcommand{\bR}{\bm R}

\begin{document}
\title{Production of exotic tetraquarks $QQ\bar{q}\bar{q}$ in heavy-ion collisions at the LHC}
\author{C.~E.~Fontoura}
\email{ce.fontoura@unesp.br}
\affiliation{Instituto Tecnol\'ogico de Aeron\'autica, DCTA, 12228-900 S\~ao
Jos\'e dos Campos, SP, Brazil}
\author{G.~Krein}
\email{gastao.krein@unesp.br}
\affiliation{Instituto de F\'{\i}sica Te\'orica, Universidade Estadual Paulista,
Rua Dr. Bento Teobaldo Ferraz, 271 - Bloco II, 01140-070, 
S\~ao Paulo, SP, Brazil}
\author{A.~Valcarce}
\email{valcarce@usal.es}
\affiliation{Departamento de F\'\i sica Fundamental and IUFFyM,
Universidad de Salamanca, 37008 Salamanca, Spain}
\author{J.~Vijande}
\email{javier.vijande@uv.es}
\affiliation{Unidad Mixta de Investigaci\'on en Radiof\'\i sica e Instrumentaci\'on Nuclear 
en Medicina (IRIMED), Instituto de Investigaci\'on Sanitaria La Fe (IIS-La Fe)-Universitat 
de Valencia (UV) and IFIC (UV-CSIC), 46100 Valencia, Spain}
%
%\date{\emph{Version of }\today}
%
\begin{abstract}
We investigate the production of exotic tetraquarks, $QQ\bar{q}\bar{q} \equiv T_{QQ}$ ($Q=c$ or $b$ 
and $q=u$ or $d$), in relativistic heavy-ion collisions using the 
quark coalescence model. The $T_{QQ}$ yield is 
given by the overlap of the density matrix of the constituents in the 
emission source with the Wigner function of the produced tetraquark. The tetraquark 
wave function is obtained from exact solutions of the four-body problem using realistic 
constituent models. The production yields are typically one order of magnitude
smaller than previous estimations based on simplified wave functions for the tetraquarks. 
We also evaluate the consequences of the partial restoration of chiral symmetry at the 
hadronization temperature on the coalescence probability. 
Such effects, in addition to increasing the stability of the tetraquarks, lead to an enhancement 
of the production yields, pointing towards an excellent discovery potential in forthcoming experiments. 
We discuss further consequences of our findings for the search of exotic tetraquarks in central Pb+Pb 
collisions at the LHC.
\end{abstract}
%\pacs{14.40.Lb,12.39.Pn,12.40.-y}
\maketitle
%
%%%%%%%%%%%%%%%%%%%%%%%%%%%%%%%%%%%%%%%%%%%%%%%%%%%%%%%%%%%%%%%%%
%
\section{Introduction}
\label{sec:intro}

There is a long-standing prediction that flavor-exotic four-quark states with two units of heavy 
flavor, $QQ\bar q \bar q \equiv T_{QQ}$ ($Q=c$ or $b$ and $q=u$ or $d$), are stable against decay 
into two $Q\bar q$ mesons, the binding energy increasing with the heavy-to-light 
quark-mass ratio $M_Q/m_q$~\cite{Ade82,Zouzou:1986qh,Man93}. The critical value of $M_Q/m_q$ for 
binding is somewhat model dependent, but there is nowadays a broad theoretical consensus in the 
literature {\textemdash}see Ref.~\cite{Ric18} for a recent compendium{\textemdash} about the 
existence of a deeply-bound doubly-bottom tetraquark, $T_{bb}$, with quantum numbers $(I)J^P=(0)1^+$, 
strong- and electromagnetic-interaction stable with a binding energy that might be as large 
as 100~MeV or more~\cite{Fra17,Kar17,Eic17,Karliner:2014gca,Bic16,Vij09,Ric18,Luo17,Duc13,Cza18}.
This exciting perspective is further reinforced by recent calculations predicting the stability of 
tetraquarks with distinguishable heavy quarks, 
$QQ^\prime\bar q\bar q \equiv T_{QQ^\prime}$~\cite{Kar17,Caramees:2018oue,Fra18}. 

Let us review the different recent theoretical studies leading to the stability of the $T_{bb}$ 
tetraquark. A novel lattice QCD calculation~\cite{Fra17} 
employing a nonrelativistic formulation to simulate the bottom quark finds unambiguous signals 
for a strong-interaction-stable $(0)1^+$ tetraquark, 189(10)\,MeV below the corresponding
two-meson threshold, $\bar B\bar B^*$ . With such binding, the tetraquark will be stable 
also with respect to electromagnetic decays. Ref.~\cite{Kar17} uses the mass of the 
doubly-charmed baryon $\Xi_{cc}^{++}$ recently discovered by the LHCb Collaboration~\cite{Aai17} 
to calibrate the binding energy of a $QQ$ diquark. Assuming that 
the $bb$ diquark binding energy in a $T_{bb}$ is the same as that of the $cc$ diquark in the $\Xi_{cc}^{++}$,
the mass of the $(0)1^+$  doubly-bottom tetraquark is estimated to be 215~MeV below 
the strong decay threshold $\bar B\bar B^*$. Combining heavy-quark-symmetry (HQS) mass relations of 
heavy-light and doubly-heavy-light mesons and baryons with leading-order corrections for
finite heavy-quark mass, corresponding to hyperfine spin-dependent terms and kinetic energy 
shift that depends only on the light degrees of freedom, Ref.~\cite{Eic17} predicts that the
$T_{bb}$ state is stable against strong decays. More specifically, using as input 
the masses of the doubly-bottom baryons (not yet experimentally measured) obtained by
the model calculations of Ref.~\cite{Karliner:2014gca}, Ref.~\cite{Eic17} finds 
an axial-vector tetraquark bound by 121~MeV. Ref.~\cite{Bic16} solves the Schr\"odinger 
equation with a potential extracted from a lattice QCD calculation for static heavy quarks. 
Using pion masses of $m_\pi\sim 340$\,MeV they find evidence for an isoscalar 
doubly-bottom axial-vector stable tetraquark.

When extrapolated to physical pion masses it has a binding energy of 
$90^{+43}_{-36}$\,MeV. The robustness of these predictions comes reinforced by detailed 
few-body calculations using phenomenological constituent models based 
on quark-quark Cornell-like interactions~\cite{Vij09,Ric18}, which predict that the isoscalar axial-vector 
doubly-bottom tetraquark is strong- and electromagnetic-interaction stable with a binding energy 
ranging between 144--214\,MeV for different realistic quark-quark potentials. Recent studies using 
a simple color-magnetic model come to similar conclusions~\cite{Luo17}. The QCD sum rule analysis 
of Ref.~\cite{Duc13} also points to the possibility of a stable doubly-bottom isoscalar axial-vector 
tetraquark. Finally, the recent phenomenological analysis of Ref.~\cite{Cza18} also presents
evidence in favor of the existence of a stable $T_{bb}$ state. In summary, the theoretical evidence 
seems to be very compelling and entices one to claim that the $T_{bb}$ tetraquark is 
an {\em unavoidable hadron}\,! On the other hand, the theoretical evidence for the 
existence of a $T_{cc}$ tetraquark is not so convincing~\cite{Jan04,Ric18}, the results 
depending on the dynamical model.

Tetraquarks have the simplest multiquark configuration among the exotic states reported by 
experiments up to now~\cite{Shepherd:2016dni}. The tetraquark picture was first introduced 
in the light-quark sector~\cite{Jaffe:1976ig} as an attempt to explain the inverted mass 
spectrum (inverted in comparison to the simple quark-antiquark structure favored by 
the naive quark model)
exhibited by the low-lying scalar mesons: $a_{0}(980), f_{0}(980), 
f_0(500)$ and $K^*_0(800)$~\cite{PDG18}. For the heavy tetraquarks, all observed candidates fit to 
the substructure $Q\bar{Q} q\bar{q}$ (see Refs.~\cite{Che16} for a recent compendium). However, there are not yet heavy 
tetraquarks with a $QQ\bar{q}\bar{q}$ configuration reported by experiment.
If such exotic states do exist, producing and identifying them is an extraordinary 
experimental challenge. Most of the discoveries of exotic hadrons in the last decade were made  
in $e^+e^{-}$ collisions, starting with the charmonium-like state $X(3872)$ observed
in $B\rightarrow K \, \pi^{\pm} \psi^\prime$ decays by the Belle 
Collaboration~\cite{Choi:2007wga}. In recent years, proton-proton collisions at the LHC have shown an enormous 
potential by confirming some earlier discoveries and also revealing new states. Note that a major 
difficulty in the production of a $T_{QQ}$ state in $e^+ e^-$ collisions is that two heavy-quark pairs, $Q\bar Q$, 
produced in hard scatterings must rearrange into $QQ$ and $\bar{Q}\bar{Q}$ diquarks, which makes
it a much rarer event than the production of hadrons with $Q\bar{Q}$ content~\cite{Eic17}. Despite these difficulties, 
the recent estimates in Ref.~\cite{Ali18} for the production cross sections of $T_{bb}$ and 
$T_{bc}$ tetraquarks based on Monte Carlo event generators point towards an excellent discovery 
potential in ongoing and forthcoming proton-proton collisions at the LHC. 

An alternative that circumvents those rare rearrangement processes is the production 
of quarks by coalescence in the environment of the matter produced in heavy-ion collisions 
at ultra-relativistic energies, the quark-gluon plasma (QGP), 
since the number of heavy quarks available for producing such structures is 
appreciable~\cite{{Chen:2007zp},{Cho:2011ew},{Cho:2010db}}. Along with 
the large array of applications offered by relativistic heavy-ion collisions~\cite{Foka:2016vta}, 
search of exotic hadrons in the QGP is an exciting new direction in our quest to understanding their 
structure. In generic terms, the coalescence model is based on an adiabatic approximation, 
in which the probability for the  production of, for example, a tetraquark from deconfined quarks 
is given by the overlap of the density matrix of the quark distribution with the Wigner function 
of the tetraquark. Ref.~\cite{Fries:2008hs} gives 
a review on applications of the model to hadron formation from a QGP, in which the underlying 
assumptions of the model and its successes in reproducing hadron yields in relativistic heavy-ion 
collisions are thoroughly discussed. 

Within this perspective, in this work we intend to study the production by coalescence 
of $T_{QQ}$ tetraquarks in central ${\rm Pb} + {\rm Pb}$ collisions at the LHC at $\sqrt{s_{NN}} = 2.76$~TeV 
and $\sqrt{s_{NN}} = 5.02$~TeV {\textemdash}$\sqrt{s_{NN}}$ is the total
collision energy per nucleon-nucleon pair in the center of mass (c.m.) frame.
We employ the dynamical coalescence model extensively used in exotic-hadron production reviewed recently 
in Ref.~\cite{Cho:2017dcy}. Previous studies invariably make use of a single Gaussian for the hadron 
wave functions which, although serving to obtain simple expressions for the yields, 
are far from being realistic and might be an important source of uncertainty. 
We avoid such approximation and calculate the Wigner function of the tetraquark employing the 
four-body wave function obtained from constituent models that 
correctly reproduce the low-lying meson and baryon spectra. In particular,
we use the chiral constituent quark model, $\chi$CQM, of Ref.~\cite{Val05}
and the Cornell-like interaction, AL1, of Ref.~\cite{Sem94}. Both, the $\chi$CQM and 
the AL1 models, predict a binding energy for the $T_{bb}$ tetraquark~\cite{Vij09,Ric18}
comparable to the recent HQS and Lattice QCD estimates~\cite{Fra17,Eic17,Kar17,Bic16}.

We also address the important question of how a partial restoration of chiral symmetry 
affects the coalescence process~\cite{Fries:2008hs}. Since the coalescence 
happens at nonzero temperature, at which the coalescing (light) constituent quarks have properties
different from those in vacuum, the tetraquark wave function is expected to be modified. 
The importance of such effects has already been investigated in transport~\cite{Rehberg:1995kh} and
molecular-dynamics~\cite{Marty:2012vs} descriptions of hadron production. In the present context,
this issue becomes particularly relevant for the stability of the produced tetraquark 
against two-meson decays, as not only the tetraquark mass is changed from its vacuum value, 
but the threshold energy, which is given by the sum of the masses of two mesons, 
is also modified. The in-medium stability of a $T_{QQ}$ state is of central importance for assessing 
the effects of the interactions of the tetraquark with other particles during 
the expansion of the system before kinetic freeze-out~\cite{Hong:2018mpk}. 

The paper is organized as follows. In Sect.~\ref{sec:coal} we briefly review the basic features of the
coalescence model. In Sect.~\ref{sec:Wigner} we discuss the structure of the tetraquark wave 
function employed in this work and its use in the computation of the Wigner function.
In Sect.~\ref{sec:results} we present and discuss our results in comparison with previous
estimates in the literature. We will concentrate our discussion 
on the most promising exotic tetraquark candidate, the isoscalar doubly-bottom axial-vector
tetraquark $T_{bb}$. Finally, in Sect.~\ref{sec:conclusion} 
we summarize the most important findings of our work.

%%%%%%%%%%%%%%%%%%%%%%%%%%%%%%%%%%%%%%%%%%%%%%%%%%%%%%%%%%%%%%%%%
%
\section{Coalescence of tetraquarks}
\label{sec:coal}

According to the coalescence model, the probability of producing tetraquark hadrons
from quarks in the medium formed in a QGP 
is given by the overlap of the Wigner function of the produced hadron with the 
phase-space distribution of the constituents in the medium. 
Here we follow the developments in Refs.~\cite{Cho:2010db,Cho:2011ew,Cho:2017dcy}, in which the coalescence model 
was employed to study the production of exotic hadrons in heavy-ion collisions. The implementation 
of the coalescence model in those references is particularly suitable for the present investigation 
since the Wigner function of the produced exotic hadrons is motivated by a 
nonrelativistic constituent quark model. Explicitly, the number of $T_{QQ}$ 
hadrons is given by
\be
N_{T_{QQ}} = g_{T_{QQ}}\left(\prod^4_{j=1} \frac {N_j}{g_j}\right) \int \frac{d\bP}{(2\pi)^3} 
 \frac{\int \left(\prod_{i=1}^{4}d\bp_i\,d\bx_i\,e^{- \bp^2_{i\perp}/2Tm_i}\right)
\rho^W_{\bP}(\bx_{1},\cdots,\bx_{4}; \bp_{1},\cdots, \bp_{4})}
{\int \prod_{i=1}^{4}d\bp_i\,d\bx_i\,e^{- \bp^2_{i\perp}/2Tm_i}}, 
\label{Nc-nr}
\ee
where $N_j$ is the total number of quarks of flavor $j$ produced in the collision and $g_j$ its degeneracy, 
$\bp_{i \perp}$ is the transverse momentum of a quark with flavor $i$, $T$ is the
hadronization temperature, and $\rho^W_{\bP}(\bx_{1},\cdots,\bx_{4};\bp_{1},\cdots, \bp_{4})$ 
is the Wigner function of the tetraquark. $\bP = \bp_1 + \bp_2 + \bp_3 + \bp_4$ is the c.m. 
momentum. Finally $g_{T_{QQ}}$ is the degeneracy factor of the tetraquark given by 
$(2 J_{T_{QQ}} + 1)(2 I_{T_{QQ}} + 1)$~\footnote{ We will be interested in an isoscalar
axial-vector tetraquark, thus $J_{T_{QQ}} = 1$ and $I_{T_{QQ}} = 0$. The de\-ge\-neracy factor
$g_j$ of quarks of flavor $j$ would correspond to its color-spin dege\-ne\-racy, i.e., $(3 \times 2)$
for each constituent~\cite{Chen:2007zp,Lee:2013nya}.}.

Ref.~\cite{Cho:2017dcy} reviews the details and discusses the different hypotheses
made in arriving to the expression of $N_{T_{QQ}}$. The most important ones are: neglect of 
transverse flow of the produced matter, consideration of only the central unit rapidity assuming 
uniform rapidity quark distributions, use of nonrelativistic approximations, and use of a Boltzmann 
distribution for the transverse quark momenta for the phase-space distribution of the quarks. In 
addition, it is further assumed that the time in which the coalescence occurs after the collision is 
large compared with the internal time scale of the hadron, what allows to omit the contribution 
from the longitudinal relative momenta.  
It is worth to note that Ref.~\cite{Sun:2017ooe} has derived an alternative implementation of the 
coalescence model for the study of exotic hadrons overcoming some of the approximations
mentioned above, as it could be to consider relativistic
effects or finite-size effects of the produced cluster
relative to the emission source. However, it is explicitly stated in Ref.~\cite{Sun:2017ooe}
that the alternative implementation gives significantly
different predictions for exotic hadrons with nonzero
orbital angular momentum, which is not the case of the $T_{bb}$ state
of our interest. In all other cases it gives results
very close to the original derivation of Refs.~\cite{Cho:2010db,Cho:2011ew,Cho:2017dcy}
that we follow in the present work.

The integration over $\bP$ in Eq.~(\ref{Nc-nr}) can be done by expressing
the tetraquark wave function in terms of the following Jacobi coordinates~\cite{Vijande:2009zs,
Vijande:2009ac}:
\bea
\bR &=& \frac{1}{M}\left(m_1\, \bx_1 + m_2 \,\bx_2 + m_3\, \bx_3 + m_4\, \bx_4\right), \nonumber
\\[0.1cm]
\br_{1} &=& \bx_1 - \bx_2, \nonumber
\\[0.1cm]
\br_{2} &=& \bx_3 - \bx_4, 
\\[0.1cm]
\br_{3} &=& \frac{m_1\,\bx_1 + m_2 \,\bx_2}{m_1+m_2} -\frac{m_3\, \bx_3 + m_4 \,\bx_4}{m_3+m_4}\, , \nonumber
\label{rel-coord}
\eea
where $M = \sum_i^4 m_i$, with $m_i$ being the masses of the constituent quarks. In terms
of these coordinates, the Wigner function is given by
\bea
\rho^W_{\bP}(\bx_{1}, \bx_{2},\bx_{3},\bx_{4};\bp_{1}, \bp_{2}, \bp_{3}, \bp_{4}) &=&
(2\pi)^3 \, \delta (\bP - \bp_1 - \bp_2 - \bp_3 - \bp_4) \nn\\[0.1cm]
&& \times \, \rho^W_{\rm int}(\br_{1},\br_{2},\br_{3}; \bk_1,\bk_2,\bk_3),
\label{wigner-func}
\eea
where $\rho^W_{\rm int}(\br_{1},\br_{2},\br_{3}; \bk_1,\bk_2,\bk_3)$ is given 
in terms of the tetraquark wave function $\psi(\br_1,\cdots, \br_3)$ as 
\bea
\rho^W_{\rm int}(\br_{1},\br_{2},\br_{3}; \bk_1,\bk_2,\bk_3) &=& \int \left(\prod^3_i d\br'_{i}\, 
e^{-i \bk_i \cdot r'_{i}}\right) \psi(\br_1 + \br'_1/2, \br_2 + \br'_2/2, \br_3 + \br'_3/2) \nn \\
&& \times \, \psi^*(\br_1 - \br'_1/2, \br_2 - \br'_2/2, \br_3 - \br'_3/2),
\label{wigner-jacobi}
\eea
with $\bk_i$ being the conjugate momenta relative to $\br_i$. Integrating over $\bP$
and performing the phase-space integrals in the denominator of Eq.~(\ref{Nc-nr}), one obtains
\bea
N_{T_{QQ}}  = g_{T_{QQ}} \frac{\prod^4_{j=1} \left(N_j/g_j\right)}{\prod^3_{i=1}\left[V\,(2\pi T \mu_i)\right]}\,
{\cal F}_{T_{QQ}}(T),
\label{Nc-sf}
\eea
where $V$ is the volume of the source and we have defined the temperature-dependent overlap function ${\cal F}_{T_{QQ}}(T)$:
\bea
{\cal F}_{T_{QQ}}(T) = 
\int \left(\prod^3_{i=1}d\bk_{i\perp} \, d\br_i \,e^{- \bk^2_{i\perp}/{2T\mu_{i}}}\right)  
\rho^{W}_{\rm int}(\br_{1},\br_{2},\br_{3}; \bk_1,\bk_2,\bk_3),
\label{wigner-overlap}
\eea
with the reduced masses $\mu_i$ given by 
\bea
\mu_{1} &=& \frac{m_{1}m_{2}}{m_{1}+m_{2}},
\hspace{0.5cm} \mu_{2} = \frac{m_{3}m_{4}}{m_{3}+m_{4}},
\hspace{0.5cm}
\mu_{3} = \frac{(m_{1} + m_{2})(m_{3} + m_{4})}{M}.
\label{def-reduced-mass-variables}
\eea

Note that in addition to the explicit $T$ dependence due to the presence of the Boltzmann 
distribution in Eq.~(\ref{wigner-overlap}), ${\cal F}_{T_{QQ}}(T)$ might also acquire an 
implicit $T$~dependence through the parameters of the constituent model when chiral symmetry
restoration effects are taken into account, as it is discussed in the forthcoming sections. 

%%%%%%%%%%%%%%%%%%%%%%%%%%%%%%%%%%%%%%%%%%%%%%%%%%%%%%%%%%%%%%%%%
%
\section{Evaluation of the tetraquark Wigner function}
\label{sec:Wigner}

We evaluate the tetraquark Wigner function with a four-body wave function obtained from realistic
constituent quark models by means of a generalized Gaussian variational method. As mentioned in
the introduction, we will present results for two different constituent quark models, $\chi$CQM 
and AL1, to check the robustness of our predictions. Chiral symmetry restoration effects will be
addressed by means of the $\chi$CQM model, that has already been used for various studies of hadron masses and
hadron-hadron interactions in-medium~\cite{{Carames:2016qhr},{Carames:2018xek}}.
 
Let us first of all briefly summarize the most important features of the constituent quark
models. The $\chi$CQM takes into account short-distance perturbative QCD effects through a 
one-gluon-exchange potential. In addition to the masses for the constituent quarks, 
dynamical chiral symmetry breaking generates (pseudo) Goldstone bosons, introduced 
as explicit degrees of freedom via $\pi$ and $\sigma$ fields. This aspect makes 
the model ideally suited to study the effects of partial restoration of chiral symmetry. 
Quark confinement is incorporated via an effective potential that contains 
string-breaking effects. The charm or bottom and light quarks interact only 
via one-gluon exchange and, of course, are subject to the same confining 
potential{\textemdash}for a detailed review of the model we refer the reader to Refs.~\cite{Val05}. 
The quark-quark potential in the AL1 model contains a chromoelectric part made of a Coulomb-plus-linear 
interaction together with a chromomagnetic spin-spin term described by a regularized Breit-Fermi 
interaction with a smearing parameter that depends on the reduced mass of the 
interacting quarks. Further details of the AL1 model are given in Ref.~\cite{Sem94}.

The tetraquark wave function is taken to be a sum over all allowed channels with 
well-defined symmetry properties~\cite{Vijande:2009zs,Vijande:2009ac}: 
\be
\psi(\br_{1},\br_{2},\br_{3}) = \sum^{6}_{\kappa=1} \chi^{csf}_{\kappa} \, 
R_{\kappa}(\br_{1},\br_{2},\br_{3}),
\label{trial-wave-function}
\ee
where  $\chi^{csf}_\kappa$ are orthonormalized 
color-spin-flavor vectors and $R_{\kappa}(\br_{1},\br_{2},\br_{3})$ is the radial part 
of the wave function of the $\kappa-$th channel. 
In order to get the appropriate symmetry properties
in configuration space, $R_{\kappa}(\br_{1},\br_{2},\br_{3})$ 
is expressed as the sum of four components, 
\bea
R_{\kappa}(\br_{1},\br_{2},\br_{3})&=& \sum^4_{r=1} w(\kappa,r)  \, R^r_\kappa(\br_{1},\br_{2},\br_{3}) ,
\label{Rk}
\eea
where $w(k,r) = \pm 1$. Finally, each $ R^r_\kappa(\br_{1},\br_{2},\br_{3})$ is expanded in terms 
of $n$ generalized Gaussians
\bea
R^r_\kappa(\br_{1},\br_{2},\br_{3}) = \sum^{n}_{i=1} \alpha^{\kappa}_i \,
e^{ - a^{i}_\kappa\,\br^{2}_{1} - b^{i}_\kappa \,\br^{2}_{2} - c^{i}_\kappa \,\br^{2}_{3}
- d^{i}_\kappa \, s_1(r) \,\br_{1}\cdot\br_{2} - e^{i}_\kappa \, s_2(r) \,\br_{1}\cdot\br_{3}
- f^{i}_\kappa \, s_3(r) \,\br_{2}\cdot\br_{3}} ,
\label{Rkr}
\eea
where $s_1(r),\cdots,s_3(r)$ are equal to $\pm 1$ and 
$a^{i}_\kappa,\cdots, f^{i}_\kappa$ are variational parameters. The latter are determined by 
minimizing the intrinsic energy of the tetraquark {\textemdash}see Ref.~\cite{Vijande:2009ac}
for further details about the wave function and the minimization procedure. 

The tetraquark will be stable under the strong interaction if its total energy, $E_{T_{QQ}}$, lies below 
all allowed two-meson thresholds. Thus, one can define the difference between the mass 
of the tetraquark, $E_{T_{QQ}}$, and that of the lowest two-meson threshold, $E(M_{1},M_{2})$, 
namely:
\be
\Delta E_{T_{QQ}} = E_{T_{QQ}} - E(M_{1},M_{2}),
\label{Delta-E}
\ee
where $E(M_{1},M_{2})$ is the sum of the masses of the mesons $M_1$ and $M_2$. 
When $\Delta E_{T_{QQ}} < 0$, all fall-apart decays are forbidden and, therefore, a 
strong-interaction stable state is warranted. When $\Delta E_{T_{QQ}} \geq 0$ one is simply
dealing with a state in the continuum. Another quantity of interest is the root-mean-square 
(r.m.s.) radius of the tetraquark, ${\rm RMS}_{T_{QQ}}$, given by~\cite{Vij09}:
\be
{\rm RMS}_{T_{QQ}} = \left[\frac{\sum^4_{i=1} m_i \langle (\bx_i - \bR)^2 \rangle}{\sum^4_{i=1} m_i} 
\right]^{1/2}.
\label{RMS}
\ee

The Gaussian nature of the radial functions $R^r_\kappa(\br_{1},\br_{2},\br_{3})$ allows one to 
obtain an analytical expression for the overlap function ${\cal F}_{T_{QQ}}(T)$.  
Substituting Eq.~(\ref{trial-wave-function}) into Eq.~(\ref{wigner-jacobi}), the Wigner
function $\rho^W_{\rm int}(\br_{1},\br_{2},\br_{3}\,;\,\bk_1,\bk_2,\bk_3)$ can be written as
\bea
\rho^W_{\rm int}(\br_{1},\br_{2},\br_{3}\,;\,\bk_1,\bk_2,\bk_3)&=&
\sum^6_{\kappa=1}\sum^4_{r,r'=1}w(\kappa,r) w(\kappa,r')\sum^{3}_{i,j=1} \alpha^\kappa_i \alpha^\kappa_j
\nn \\
&& \times \, 
\int \left(\prod^3_{i=1} d\br'_i \, e^{-i \bk_i\cdot \br'_i}\right)  \,e^{- {\cal E}_W(\br^\pm_1, \br^\pm_2, \br^\pm_3)}, 
\label{wigner_final}
\eea
where $ \br^\pm_i = \br_i \pm \br'_i/2$ and 
\bea
{\cal E}_W(\br^\pm_1, \br^\pm_2, \br^\pm_3) &=&
a^{i}_k (\br^+_1)^2 
+ b^{i}_k (\br^+_2)^2
+ c^{i}_k(\br^+_3)^2
+ a^{j}_k (\br^-_1)^2
+ b^{j}_k (\br^-_2)^2
+ c^{j}_k (\br^-_3)^2
\nn\\[0.1cm]
&&
+ d^{i}_k s_1(r) \br^+_1\cdot \br^+_2
+ e^{i}_k s_2(r) \br^+_1\cdot \br^+_3
+ f^{i}_k s_3(r) \br^+_2\cdot \br^+_3
+ d^{j}_k s_1(r) \br^-_1\cdot \br^-_2
\nn\\[0.1cm]
&&
+ e^{j}_k s_2(r) \br^-_1\cdot \br^-_3
+ f^{j}_k s_3(r) \br^-_2\cdot \br^-_3
\label{EW} \, .
\eea
The overlap function ${\cal F}_{T_{QQ}}(T)$ is obtained by performing the eight-dimensional integral 
over the variables $\br_i$, $\br'_i$ and $\bk_{i\perp}$. The integrals can be done analytically, most easily 
using Cartesian coordinates, since all of them are of the form, 
\be
\int_{-\infty}^{\infty} d\xi \, e^{-a\,\xi^{2} \pm b \xi } = \left(\frac \pi a \right)^{1/2}\,
e^{b^{2}/4a},
\ee
with $a$ real and $b$ real or complex.

As we have discussed in the introduction, we will concentrate our discussion 
on the most promising exotic tetraquark candidate, the isoscalar doubly-bottom axial-vector
tetraquark $T_{bb}$, about whose existence there exists a broad theoretical 
agreement~\cite{Fra17,Kar17,Eic17,Karliner:2014gca,Bic16,Vij09,Ric18,Luo17,Duc13,Cza18}.
In the lowest-lying tetraquark configuration all four-quarks are in a relative $S$ wave. 
Thus, the tetraquark shows a separate dynamics for the compact heavy quark
in a color antitriplet (see Fig. 8 of Ref.~\cite{Ric18} and Table II of
Ref.~\cite{Vijande:2009zs}), and therefore due to Fermi statistics spin 1,
and the light antiquarks bound to a color triplet to obtain a total color singlet.
To satisfy the Pauli principle, the flavor-antisymmetric light-antiquark pair
must have spin 0 while the flavor-symmetric has spin 1. The one-gluon exchange
is much more attractive for the {\it good} antidiquark, a color triplet with spin and
isospin 0. Thus, the total spin and parity of the {\em unavoidable} $T_{bb}$ tetraquark
are $J^P=1^+$ and its isospin would $I=0$~\cite{Eic17,Kar17}.

%%%%%%%%%%%%%%%%%%%%%%%%%%%%%%%%%%%%%%%%%%%%%%%%%%%%%%%%
%
\section{Results}
\label{sec:results}

We present results for $N_{T_{QQ}}$ obtained with the $\chi$CQM and the AL1 models. 
We also investigate the effects of a finite-temperature partial
chiral symmetry restoration on $N_{T_{QQ}}$ using the $\chi$CQM. Finite temperature effects are
incorporated in those parameters of the model related to the dynamical breaking of chiral symmetry, 
namely the masses of the constituent quarks and of the $\sigma$ and $\pi$ mesons, and the couplings
of the light constituent quarks to that mesons. For the temperature dependence of 
those parameters, we use predictions of the Nambu--Jona-Lasinio model~\cite{{Nam61}}, following
the strategy set up in our work in Ref.~\cite{Carames:2016qhr}, in which the effects of a hot 
and dense medium on the binding energy of hadronic molecules with open charm
mesons were studied. The parameters of the $\chi$CQM model are listed in Table~I of 
Ref.~\cite{Carames:2018xek}{\textemdash}the mass of the $b$ quark, not listed in that table, 
being $m_{b}=5100$~MeV. The parameters of the AL1 model have been recently summarized 
in Eq.~(32) of Ref.~\cite{Ric18}.

First, we analyze the impact of a finite-temperature partial chiral symmetry restoration on
the properties of the tetraquarks. Although only the hadronization temperature is 
of relevance for the coalescence study, it is nevertheless insightful to explore the effects of a 
partial chiral symmetry restoration as a function of $T$. For this purpose, we have selected 
four representative values close to the hadronization temperature
adequate for Pb+Pb collisions at the LHC energies of $\sqrt{s_{NN}} = 2.76$~TeV and 
$5.02$~TeV~\cite{Cho:2017dcy}: $T=100$~MeV, 
$T=120$~MeV, $T=140$~MeV, and $T=156$~MeV. The temperature dependence of quark and meson masses 
and quark-meson couplings are shown in Fig.~1 of Ref.~\cite{Carames:2016qhr}{\textemdash}for orientation,
we mention that at the highest temperature, while the pion mass is essentially the 
same as in vacuum, because it is protected by chiral symmetry, the masses of the light constituent 
quarks and of the $\sigma$ meson drop 30\% with respect to their vacuum values.

\begin{table}[t]
\caption{Tetraquark masses, $E_{T_{QQ}}$, r.m.s. radii, ${\rm RMS}_{T_{QQ}}$, and binding energies, 
$\Delta E_{T_{QQ}}$, of the $(I)J^P=(0)1^+$ $T_{bb}$ and $T_{cc}$ 
states. Energies and temperatures are listed in MeV and the r.m.s. radii in~fm.}
\begin{ruledtabular}
\begin{tabular}{lccccc}
&\multicolumn{1}{c}{Vacuum }
&\multicolumn{3}{c}{{\hspace{2.5cm}}In-medium}&   \\        
\cline{2-2}\cline{3-6}
                              & $T=$0      &$T=$100      &$T=$120      &$T=$140      &$T=$156 \\
\hline
$E_{T_{bb}}$                  & 10410   & 10402   & 10395   & 10378    & 10356    \\
${\rm RMS}_{T_{bb}}$          & 0.22    & 0.22    & 0.22    & 0.21     & 0.21     \\
$\Delta E_{T_{bb}}$           & $-$202  & $-$216  & $-$224  & $-$276   & $-$348   \\
\hline 
$E_{T_{cc}} $                 &3877     & 3870    & 3864    & 3846     & 3824     \\
${\rm RMS}_{T_{cc}}$          & 0.35    & 0.35    & 0.34    & 0.34     & 0.34    \\
$\Delta E_{T_{cc}}$           & $-$60   & $-$70   & $-$76   & $-$120   & $-$180 \\
\end{tabular}
\end{ruledtabular}
\label{Tab1}
\end{table}

Table~\ref{Tab1} displays results for the masses, $E_{T_{QQ}}$,
r.m.s. radii, ${\rm RMS}_{T_{QQ}}$, and the binding energies, 
$\Delta E_{T_{QQ}}$, of the $(I)J^P=(0)1^+$ $T_{bb}$ and $T_{cc}$
states in vacuum and for the selected temperatures. We note that the lowest two-meson threshold 
for $T_{bb}$ and $T_{cc}$ corresponds to ${\bar B}{\bar B}^*$ and
$D D^*$ in relative $S-$wave, respectively. It can be seen that the tetraquarks are compact 
structures instead of molecular ones, with r.m.s. radii much smaller than $1$~fm that 
remain almost constant with $T$. With respect to their binding energies, the temperature 
affects the stability of the tetraquarks making them more stable as~$T$ increases. This is mainly due to
a larger threshold energy, as can be inferred from Fig.~2(a) of Ref.~\cite{Carames:2016qhr},
where the temperature dependence of the meson masses has been evaluated.
This latter feature is very important: even 
when the tetraquark masses are almost $T-$independent, they become more stable as $T$ increases, indicating that 
chiral symmetry restoration has a larger impact on the masses of $D$ and $\bar B$ mesons than 
on the tetraquarks, $T_{QQ}$. Clearly, the improved stability of the tetraquarks at finite~$T$ 
is a welcome feature for their formation in the matter produced in a heavy-ion collision; because, as 
we discuss further ahead, once they have been formed,
the probability to be destroyed by subsequent interactions with other hadrons 
(mainly pions) of the medium is diminished.

\begin{table}[t]
\caption{Number of $b$ and $c$ quarks per unit rapidity at midrapidity 
in 0$-$10\% central collision at RHIC and LHC taken from Ref.~\cite{Cho:2017dcy}. 
In the last two columns, under Extrapolation, we give the estimates for $N_b$ and $N_c$ 
at higher energies, obtained from a linear extrapolation of the data at the three 
lower energies.} 
\begin{ruledtabular}
\begin{tabular}{ccccccc}
& \multicolumn{1}{c}{RHIC} &\multicolumn{2}{c}{LHC} &\multicolumn{2}{c}{Extrapolation} \\
\cline{2-2} \cline{3-4} \cline{5-6} 
&0.2~TeV&2.76~TeV&5.02~TeV&10~TeV&15~TeV\\
\hline
$N_{b}$ & 0.031  & 0.44  & 0.71 & 1.43 & 2.14\\
$N_{c}$ & 4.1    & 11    & 14   & 25   & 35  
\end{tabular}
\end{ruledtabular}
\label{Tab2}
\end{table}

Next, we present results for the tetraquark yields, $N_{T_{QQ}}$, given by Eq.~(\ref{Nc-sf}).  
For this purpose, it is necessary to specify the values of the hadronization temperature~$T$, the volume $V$, 
and the quark numbers $N_{i}$\,($i=q,b,c$) (the latter being understood as being per unit 
of rapidity at midrapidity). We use the values given in Ref.~\cite{Cho:2017dcy} which are suitable 
for Pb+Pb collisions at the LHC energies of $\sqrt{s_{NN}} = 2.76$~TeV and $5.02$~TeV: 
$T=156$~MeV, $V=5380$~fm$^{3}$, $N_{q}=700$, and $N_b$ and $N_c$ are given in Table~\ref{Tab2}.
Note that the heavy quarks are produced by hard scatterings at the early stage of the collisions 
and as such are $\sqrt{s_{NN}}-$dependent. For simplicity, for the RHIC energy of $\sqrt{s_{NN}} = 0.2$~TeV,
we use the same values for $T$ and $V$. In addition, to assess a possible enhancement in the number of 
produced tetraquarks by an increase of the collision energy $\sqrt{s_{NN}}$, we have performed a 
linear extrapolation of the RHIC and LHC data on $N_b$ and $N_c$ to $\sqrt{s_{NN}}=10$~TeV and 15~TeV. 
The results of the extrapolation are shown in the last two columns of Table~\ref{Tab2}.
It is important to emphasize that in a comparison of the production cross section of $T_{bb}$ states 
to that of doubly-bottom baryons, Ref.~\cite{Ali18} finds that the latter is 
2.4 the former. This result represents an excellent discovery potential of $T_{bb}$ tetraquarks
in the near future in a dedicated search at the LCH at $\sqrt{s_{NN}}$ = 13 TeV, and points to
the necessity of obtaining predictions for the yields in that range of energies.
 
\begin{table}[t]
\caption{Tetraquark yields for central Pb+Pb collisions at LHC energies $\sqrt{s_{NN}} = 2.76$~TeV 
and $\sqrt{s_{NN}} = 5.02$~TeV. Results in the two columns under Extrapolation are the estimates
at higher energies using for $N_b$ and $N_c$ the corresponding values shown in Table~\ref{Tab2}. 
The temperature used in the Boltzmann distribution and for the chiral symmetry restoration 
effects is $T=156$~MeV.}
\begin{ruledtabular}
\begin{tabular}{cccccc}
&\multicolumn{2}{c}{LHC} &\multicolumn{2}{c}{Extrapolation} \\
\cline{2-3} \cline{4-5} 
&2.76~TeV&5.02~TeV& 10~TeV&15~TeV\\
%\hline
%
&\multicolumn{4}{c}{\underline{No chiral restoration}} \\
%&\multicolumn{2}{c}{Extrapolation} \\
%\cline{2-3} \cline{4-5} 
%\\ 
%\hline
$N_{T_{bb}}$ & $6.2 \times 10^{-9}$  & $1.6 \times 10^{-8}$ 
& $6.6 \times 10^{-8}$ & $1.5 \times 10^{-7}$\\
$N_{T_{cc}}$ & $2.4 \times 10^{-5}$  & $3.8\times 10^{-5}$ 
& $1.2 \times 10^{-4}$  & $2.4 \times 10^{-4}$  \\
%\hline
&\multicolumn{4}{c}{\underline{Chiral restoration}} \\
$N_{T_{bb}}$ & $1.3 \times 10^{-8}$  & $3.4 \times 10^{-8}$ 
& $1.4 \times 10^{-7}$ 
& $3.1 \times 10^{-7}$ \\
$N_{T_{cc}}$ & $4.0 \times 10^{-5}$  & $6.5 \times 10^{-5}$  
& $2.1 \times 10^{-4}$  
& $4.1 \times 10^{-4}$
\end{tabular}
\end{ruledtabular}
\label{Tab3}
\end{table}

Table~\ref{Tab3} displays our predictions for the yields of the  $T_{bb}$ and $T_{cc}$ tetraquarks,
$N_{T_{bb}}$ and  $N_{T_{cc}}$, for central Pb+Pb collisions at the LHC energies of $\sqrt{s_{NN}} = 2.76$~TeV  
and $\sqrt{s_{NN}} = 5.02$~TeV. Also shown are the predictions for the higher extrapolated energies.  
We consider first the situation when chiral symmetry restoration effects on the coalescence are
ignored{\textemdash}the corresponding results appear in the columns under ``No chiral restoration''. 
Let us compare the results with previous studies based on the coalescence model. 
Ref.~\cite{Cho:2011ew} presents predictions for $T_{cc}$ considering the
situations that the $T_{cc}$ is either a molecular state or a compact multiquark.
In the first situation hadron coalescence is employed and for the latter quark coalescence. Different
temperatures are used in each case, kinetic freeze-out temperature of 125~MeV for hadron coalescence and 
hadronization temperature of 175~MeV for quark coalescence. That reference employs a single Gaussian 
to represent the hadron and molecular wave functions, with different width parameters, of course. 
The results are in the range $2.4\times10^{-5}$ 
(hadron coalescence) to $4.0\times10^{-5}$ (quark coalescence) for 
$\sqrt{s_{NN}}$=0.2~TeV and $4.1\times 10^{-4}$ (hadron coalescence) to $6.6\times 10^{-4}$ (quark coalescence)   
for $\sqrt{s_{NN}}$=5.5 TeV. The latter might be compared with our result, $3.8\times 10^{-5}$ 
for $\sqrt{s_{NN}}$=5.02 TeV and $T=156$~MeV, which is almost one order of magnitude smaller. 
As has been discussed above, 
Ref.~\cite{Hong:2018mpk} finds essentially the same numbers of Ref.~\cite{Cho:2011ew}. 
In all cases, it is clear that the yield from the coalescence model for compact multiquark states is smaller than
that for the usual quark configurations as a results of the suppression owing to the coalescence
of additional quarks.

We have also obtained results with the AL1 quark model. That model predicts a compact $T_{bb}$ 
bound state with the same r.m.s. radius predicted by the $\chi$CQM, 0.22~fm, although 
with a binding energy 30\% smaller. For the tetraquark yields, the model predicts: $N_{T_{bb}} = 8.8 \times
10^{-9}$ for $\sqrt{s_{NN}} = 2.76$~TeV, and $N_{T_{bb}} = 3.3 \times 10^{-8}$ for $\sqrt{s_{NN}} = 5.02$~TeV.
Although some very small differences can be observed between the yields predicted by the 
two realistic models, $\chi$CQM and AL1, the order of magnitude is the same, being smaller than the
simplistic approximation of considering a single Gaussian for the tetraquark wave function. 
It is reassuring that the results are stable with respect to the difference in the 
binding energy

When including effects due to partial chiral symmetry restoration, Table~\ref{Tab3} reveals that the 
yields increase roughly by a factor two. Such a modest influence is a consequence of the small effect 
of partial chiral symmetry restoration on the r.m.s. radii. Although those effects also 
modify the masses $m_q$ entering the Boltzmann distribution, they essentially cancel out
in the expression for $N_{T_{QQ}}$, see Eq.~(\ref{Nc-nr}). However, as already mentioned, the important
feature of the partial chiral symmetry restoration is the improved stability of the tetraquark 
in-medium, due to the larger threshold for two-meson decays. 

The role played by hadronic effects, i.e.,
changes occurred in the production rate due to the interaction with other particles during the 
expansion of the medium, was discussed for the case of $T_{cc}$ states in Ref.~\cite{Hong:2018mpk}. 
The authors conclude that these hadronic effects 
are negligible for the case of compact states. We recall that the $T_{bb}$ wave function, Eq.~(\ref{trial-wave-function}),
used in the present work is composed dominantly by color configurations that are a color-triplet
for the light quarks and a color antitriplet for the bottom quarks (see the penultimate column in
Table II of Ref.~\cite{Vijande:2009zs} and Fig. 8 of Ref.~\cite{Ric18}) which can be decomposed 
into $b\bar q$ color-singlet $\bar B_1 \bar B^*_1$ and color-octect $\bar B_8 \bar B^*_8$ 
states. This corresponds to a compact state with a large hidden-color component
that, differently from $Q\bar Q q\bar q$ states,
cannot be expressed in terms of a single two-meson state{\textemdash}see Ref.~\cite{Vijande:2009ac} 
for detailed discussions. Therefore, one can safely assume that the abundance of $T_{bb}$ calculated 
at the QGP phase will not change significantly during the expansion of the hadronic matter as a 
result of absorption by other hadrons in the medium. 
The signal for the formation of such states would be through the detection
of their weak-decay products with several Cabibbo allowed two- and three-body 
decay channels~\cite{Eic17,Kar17}:
$T_{bb}^- \rightarrow \Xi^0_{bc} \, \bar p$,
$T_{bb}^- \rightarrow B^- \, D^+ \, \pi^-$, or
$T_{bb}^- \rightarrow B^- \, D^+ \, \ell^- \, \bar \nu_\ell$,
that offer enormous discovery potential as they do not contain identical quarks
or antiquarks, which will induce a spin-statistic suppression. Recent flavor $SU(3)$ 
relations based on a chromomagnetic model~\cite{Xing:2018bqt} confirm the adequacy
of these channels to search for doubly heavy tetraquark states at the LHCb and 
Belle II experiments. 

%%%%%%%%%%%%%%%%%%%%%%%%%%%%%%%%%%%%%%%%%%%%%%%%%%%%%
%
\section{Conclusions and perspectives}
\label{sec:conclusion}

The question on whether $QQ\bar{q}\bar{q} \equiv T_{QQ}$ hadrons can be experimentally observed is 
of great contemporary interest. Observation of such states will be of help in our quest to understand 
the structure of the newly observed states with quark compositions beyond the traditional quark-antiquark 
and three-quark configurations. A major challenge in such a program 
is the lack of  experimental information 
on the production of these exotic hadronic systems in-vacuum and in-medium. With this motivation, 
we have investigated the production of $T_{bb}$ and $T_{cc}$ tetraquarks in relativistic heavy-ion 
collisions in central Pb+Pb collisions at the LHC in the framework of the quark coalescence model. 
To our knowledge, this is the first study in which a four-quark wave function 
obtained by solving exactly the four-body problem using realistic constituent models was used to calculate
the hadron Wigner function. In addition, the effects of a partial restoration of 
chiral symmetry on the coalescence probability have been investigated. We have found that the order of magnitude of the 
predictions for the tetraquark yields is not modified when using different realistic
constituent models, either the $\chi$CQM or the AL1. However,  
the obtained production yields are typically one order of magnitude
smaller than previous estimations based on simplified wave functions for the tetraquarks. 

Our results indicate that the $(I)(J)^P = (0)1^+$ $T_{bb}$ tetraquark is a compact state.
It becomes more stable in-medium, when effects of a partial restoration of chiral symmetry are
taken into account, leading to an increase in the production yields by a factor 
roughly equal to two. The improved in-medium stability
of the tetraquarks implies a smaller probability to be destroyed by 
subsequent interactions with other hadrons (mainly pions) of the medium.
Therefore, the number of produced tetraquarks is given essentially by that 
calculated at the QGP phase, which essentially depends on the structure of the
state. In short, our results also suggest that measuring the $T_{bb}$ tetraquark
from heavy-ion collisions would inform us about the nature of its structure.

%
%%%%%%%%%%%%%%%%%%%%%%%%%%%%%%%%%%%%%%%%%%%%%%%%%%%%%
%
\section{acknowledgments}
Useful discussions with Marina Nielsen are gratefully acknowledged. This work has been partially funded
by Ministerio de Econom\'\i a, Industria y Competitividad and EU FEDER under Contract No. FPA2016-77177
and by a bilateral agreement Universidad de 
Salamanca - Funda\c{c}\~ao de Amparo \`a Pesquisa do Estado de S\~ao Paulo - FAPESP Grant 
No. 2017/50269-7. Partial financial support is also acknowledged from Conselho 
Nacional de Desenvolvimento Cient\'{\i}fico e Tecnol\'ogico - CNPq, 
Grants No. 168445/2017-4 (C.E.F.), 305894/2009-9 (G.K.), 464898/2014-5(G.K) (INCT F\'{\i}sica 
Nuclear e Apli\-ca\-\c{c}\~oes), and Funda\c{c}\~ao de Amparo \`a 
Pesquisa do Estado de S\~ao Paulo - FAPESP, Grant No. 2013/01907-0 (G.K.). 
%
%%%%%%%%%%%%%%%%%%%%%%%%%%%%%%%%%%%%%%%%%%%%%%%%%%%%%
%

\end{document}